\documentstyle[preprint,aps]{revtex}
\tightenlines
\begin{document}
\draft

\preprint{\vbox{\hbox{JLAB-THY-98-47}\hbox{MIT/CTP Preprint \# 2799}}}
\title{Spin-Orbit Final-State Interaction in the Framework of
Glauber Theory for $(e,e'p)$ Reactions}

\author{S. Jeschonnek$^1$ and T. W. Donnelly$^2$}

\address{\small \sl 
(1) Jefferson Lab, 12000 Jefferson Ave, Newport News, VA 23606 
(2) Center for Theoretical Physics, Laboratory for
Nuclear Science and Dept. of Physics, Massachusetts Institute of
Technology, Cambridge, MA 02139}

\date{\today}
\maketitle

\begin{abstract}
We investigate the reactions $D(e,e'p)n$ and $D(\vec e,e'p)n$ at GeV
energies and discuss the opportunities to distinguish between
different models for the nuclear ground state by measuring the
response functions.  In calculating the final-state interaction (FSI)
we employ Glauber theory, and we also include relativistic effects in
the electromagnetic current. We include not only the central FSI, but
also the spin-orbit FSI which is usually neglected in $(e,e'p)$
calculations within the Glauber framework and we show that this
contribution plays a crucial role for the fifth response function.
All of the methods developed here can be applied to any target
nucleus.

\end{abstract}

\section{Introduction}

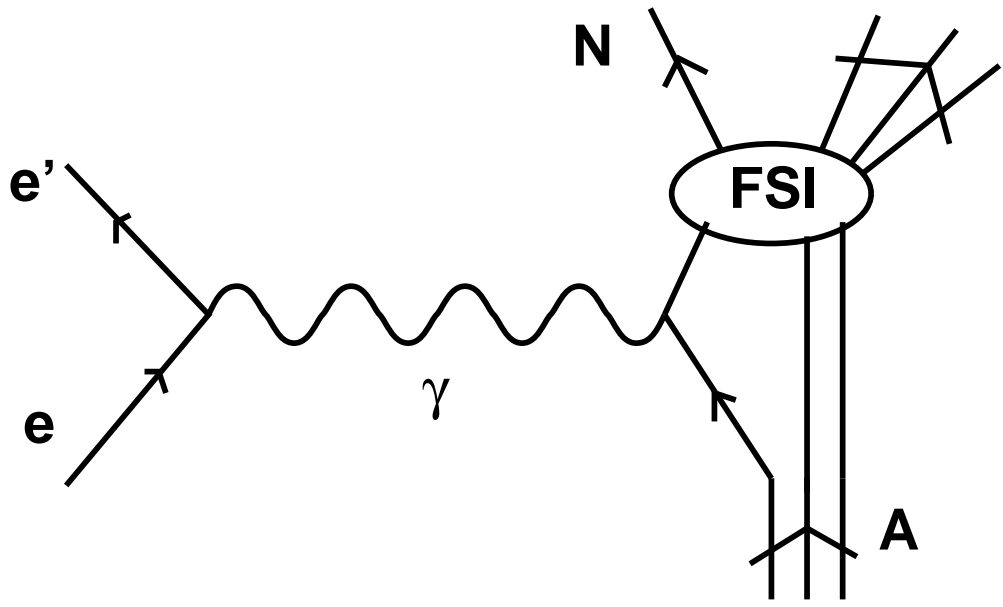
\begin{figure}
\caption{Schematic representation of the $(e,e'N)$ reaction.
The electron exchanges a photon with the nucleus, and the photon knocks out
a nucleon which undergoes f\/inal-state interactions (FSI) while leaving
the
nucleus. The scattered electron and the nucleon are detected in
coincidence.
The other hadrons in the f\/inal state remain unobserved. 
\label{figschema}}
\end{figure}

Currently, there are many exclusive electron scattering experiments
being performed or in the planning stage at medium to high energies,
e.g. at TJNAF (Jefferson Lab), Bates, NIKHEF and MAMI.  A sketch of
the typical reaction is provided in Fig.~\ref{figschema}.  In treating
the problem theoretically, ideally one would use a microscopic
Hamiltonian for the description of both the nuclear bound state and
the --- in general rather complicated --- hadronic f\/inal state. Such
a microscopic Hamiltonian would also contain a minimal coupling to the
photon, so that the initial state, f\/inal state and electromagnetic
current operator would all be treated consistently in a microscopic,
relativistic fashion. In practice, it is very dif\/ficult to perform a
consistent calculation of both nuclear ground state and f\/inal
hadronic scattering state and, especially for medium and heavy nuclei,
it is unlikely that a consistent, fully microscopic and therefore
relativistic treatment will be available in the near future.

At present, there are several approximate calculations for particular
electronuclear reactions available for few-body systems
\cite{tjon,vano,gross}. However, it is dif\/ficult to extend these
approaches for the few-body systems to energies above the pion
emission threshold. Given the considerable ef\/fort that one expects will be
put into experimental studies in this field, in the next few years
calculations in the few GeV regime for reactions at least with deuteron
targets will hopefully become available, although even this is not
trivial.  For heavy nuclei, there are only relativistic mean-field
calculations available.

Therefore, at present in practice one is forced to split the problem up 
into three
separate parts, namely the calculation of the ground state, the
calculation of the f\/inal hadronic state and the treatment of the
electromagnetic current. Although the approaches taken for those dif\/ferent
parts will not be consistent, there is a good chance that by doing the
best possible job for each separate ingredient, one will be able to
incorporate the essential physical features in the theoretical
description of exclusive electron scattering at high energies.  For
the nuclear ground-state wave function one either uses a solution of a
bound-state Schr\"odinger equation or one of the more sophisticated
microscopic few-body wave functions which are now available
\cite{vano,gross}.  
The f\/inal state is typically calculated using optical
potentials or multiple scattering theory. 
It is our aim to focus on two parts of the problem: the
electromagnetic current and the f\/inal state. The relativistic
effects in the electromagnetic current operator have been investigated
elsewhere \cite{relcur} and were found to be quite important. They are
included in all of the calculations presented in this paper.  We will discuss
methods that are applicable to all nuclei in a wide range of
energies. We develop and test these methods for the case of electron
scattering on the deuteron so that these ef\/fective methods can be
checked against the results of microscopic calculations, once the
latter are available.  

This paper is organized as follows: after this brief introduction in
Sec.~I, in Sec.~II we summarize different theoretical models for the
deuteron wave function and indicate what is expected for the Plane
Wave Impulse Approximation (PWIA). In Sec.~III, we consider the
effects of final-state interactions (FSI) and their description at
high energies by Glauber theory.
We discuss the different effects of central and spin-orbit FSI, which
are due to their different spin-structure and to their different
ranges.  Then in Sec.~IV we discuss the various possible choices of
parameters for the FSI calculation and end in Sec.~V with a brief
summary and outlook.

\section{Models for the Nuclear Ground State and the PWIA}

Before we discuss the incorporation of FSI
and the suitability of specif\/ic observables for the
discrimination between dif\/ferent theoretical models, we would like
to recall the dif\/ferences between these models. There exist several
models of the nucleon-nucleon interaction which are based on 
meson-exchange potentials \cite{bonn,paris,nijmegen} that not only make
predictions for the deuteron wave function, but also describe a
wide range of free $NN$ and meson-nucleon scattering
data. Recently, more microscopic approaches have been developed
\cite{vano,gross}.  The main dif\/ference between the meson-exchange
models lies in the dif\/ferent D-wave probability incurred: for instance, 
the Bonn model
predicts a lower D-state probability than most other models. There are
also some dif\/ferences in the very short range part ($r < 0.5$ fm)
of the S-wave.  For the later discussions, we compare results for the
Bonn wave function \cite{bonn} and the Paris wave function
\cite{paris}, as they are good representatives for the available
models and as most other theories will fall in between them.

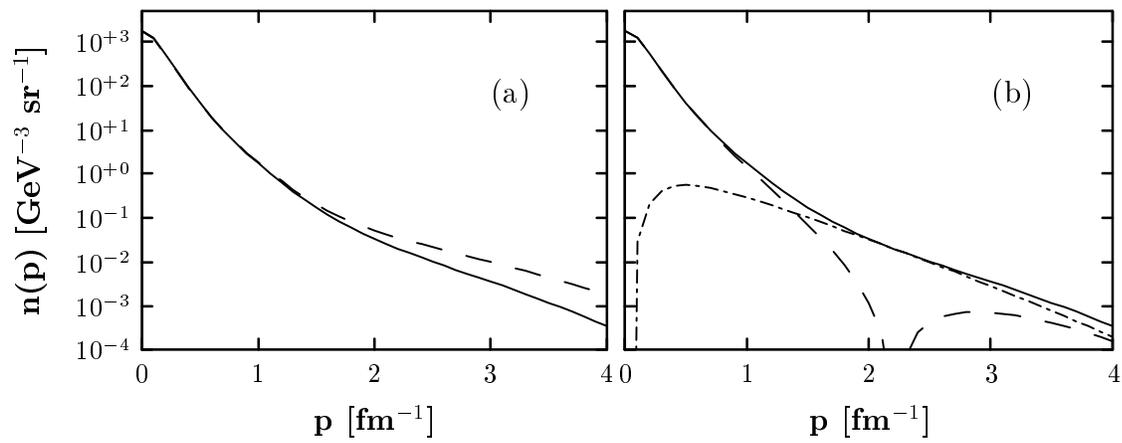
\begin{figure}
\caption{Panel (a) shows the nucleon momentum distribution $n(p)$
for the deuteron calculated with the Bonn wave function (solid line)
and the Paris wave function (dashed line). Panel (b) shows the
decomposition
of the full nucleon momentum distribution calculated with the Bonn
wave function (solid line) into the S-wave part (dashed line) and
the D-wave part (dash-dotted line). 
  \label{figmodnp}}
\end{figure}

In PWIA, the $(e,e'N)$ cross section is
proportional to the product of the electron-nucleon half-of\/f-shell cross
section times the spectral function $S(E, {\vec p})$, which gives the joint
probability to find a nucleon of energy $E$ and momentum $\vec p$ in
the nucleus:
\begin{equation}
\frac{ d^6 \sigma ^{PWIA}}{d \epsilon' d \Omega_e d \Omega_N d E_N} =
K \sigma_{eN} S(E,{\vec p}) \, ,
\label{sigpwia}
\end{equation}
where $K$ is a kinematic factor and $\sigma_{eN}$ is the 
electron-nucleon half-of\/f-shell cross section.  In the case of the deuteron, 
the energy $E$ is fixed by the binding energy and the spectral function
reduces to the momentum distribution $n(p)$ which does not depend on
the direction of the momentum $\vec p$, as long as unpolarized
deuterium is considered. Dif\/ferent predictions for the momentum
distribution $n(p)$ are shown in Fig.~\ref{figmodnp}a. It should be
noted that although the discrepancy for higher momenta is large --- up
to a factor of five --- the momentum distribution varies over seven orders
of magnitude, and the agreement between the models is quite good overall. 
It is interesting and important to glean information on the high momentum
part of $n(p)$ from experiment, as it is this kinematic region that is
expected to tell us about the transition from hadronic
degrees of freedom to quark-gluon degrees of freedom.

The dif\/ferent predictions for momenta $p > 1.5$ fm$^{-1}$ stem from the
dif\/ferent D-wave probabilities: in Fig.~\ref{figmodnp}b, the
decomposition
of the momentum distribution into S-wave and D-wave is shown, and one 
can see clearly that the momentum distribution is dominated by the 
D-wave for momenta of 1.5 fm$^{-1}$ and higher. 

Of course, the simple picture of PWIA is only an approximation, as the
outgoing nucleon will in fact interact with the residual system, and this
interaction will break the simple factorized relation of Eq.~(\ref{sigpwia})
involving the single-nucleon cross section and nuclear spectral function.  
This means that
the momentum distribution cannot be extracted from the data
immediately; in fact, it is a dif\/ficult problem to find observables
which show suf\/ficient sensitivity to the dif\/ferent models of the
ground-state wave function.  In the following, we will discuss some
possible candidates.

\section{Final-State Interaction}

Here we are interested in scattering in the GeV energy regime, and
the natural description of the FSI in this
regime is given by Glauber theory \cite{glauber}. Let us briefly
recall the change of the nature of the $NN$ interaction that occurs in
going from low to high energies. At energies of $100$ MeV or less, the
scattering of nucleons is elastic and nearly isotropic and only a few
partial waves contribute, while in contrast, at high energies, the interaction 
becomes absorptive, as new particles are produced. At an energy of $500$ MeV,
the proton-proton inelastic cross section rises sharply from values of
less than 2 mb to approximately 30 mb (for a review, see
\cite{lehar}). It then remains almost constant for energies up to
several hundred GeV.  The $NN$ interaction also becomes dif\/fractive
at high energies. The proton-nucleus elastic scattering data (see
e.g. \cite{glaumat,wallace}) taken at proton beam energies of 1
GeV and higher show a clear dif\/fractive pattern, with broad maxima
and dif\/fractive dips, reminiscent of the patterns obtained in optics
in Fraunhofer dif\/fraction. In the past, there have been great
successes in the description of proton-nucleus scattering data with
Glauber theory, see e.g. \cite{glaumat,wallace,alkhazov}, and we will
apply it here --- with the necessary modif\/ications --- to the
$(e,e'N)$ reaction.

Once FSI is
included, the generic form of the matrix elements we have to calculate
is ${\cal{M}}_{fi} = <f | S \, J_{em} | i>$, where $J_{em}$ denotes
the current operator and $S$ is the FSI operator discussed below.  An
overview of the conventions used in this paper and the relativistic
current operator employed in our calculations can be found in the
appendix.
The FSI operator $S(\vec r)$ for the 
interaction between the outgoing nucleon and a spectator nucleon reads
\begin{equation}
S(\vec r) = 1 - \theta (z) \cdot \Gamma (\vec b) \,,
\end{equation}
where the distance $\vec r$ between the two interacting nucleons
is decomposed into longitudinal and transverse parts: $\vec r = 
\vec b + z \cdot \hat q$, where $\hat q$ indicates the direction of
the virtual photon's momentum.
 The $\theta$-function indicates that the
spectator nucleon has to be in the forward hemisphere with regard to the
struck nucleon, otherwise no FSI will take place.
$\Gamma (\vec b)$ is called the prof\/ile
function of $NN$ scattering, and is related to the $NN$  scattering
amplitude $f (\vec l)$ via the Fourier transform:
\begin{equation}
\Gamma (\vec b) = \frac{1}{2 \pi i k} \int d^2 \, \vec l \, \, \exp
(-i \vec l \cdot \vec b) \, \, f (\vec l) \,.
\label{defprof}
\end{equation}
Here $\vec k$ is the incident nucleon momentum, $\vec k'$ denotes the
outgoing nucleon's momentum, and $\vec l = \vec k - \vec k'$ is the
momentum transferred in the $NN$ scattering (not to be confused with
the momentum $\vec q$ transferred to the nucleus by the electron).
The most general form for the scattering amplitude in the $N N$ center
of mass (c.m.) system assuming parity conservation, time-reversal
invariance, the Pauli principle, and isospin invariance can be written
in terms of f\/ive invariant amplitudes (see references given in
\cite{lehar}):
\begin{eqnarray}
f (\vec l) \, = & & A (\vec l) + B (\vec l) \, (\vec \sigma_1 + \vec
\sigma_2) \cdot \hat n + C (\vec l) \, (\vec \sigma_1 \cdot \hat n)
\, (\vec \sigma_2 \cdot \hat n) + \nonumber \\ & & D (\vec l) \, (\vec
\sigma_1 \cdot \hat m) \, (\vec \sigma_2 \cdot \hat m) + E (\vec l) \,
(\vec \sigma_1 \cdot \hat h) \, (\vec \sigma_2 \cdot \hat h) .
\label{eqnn}
\end{eqnarray}
The nucleon spin operators are denoted by $\vec \sigma_1$ and $\vec
\sigma_2$, and $\hat n \equiv \vec k \times \vec k' / |\vec k \times \vec
k'|$, $\hat m \equiv (\vec k - \vec k') / |\vec k - \vec k'|$, and $\hat h
\equiv (\vec k + \vec k') / |\vec k + \vec k'|$\,.

In principle, the amplitudes $A$, $B$, $C$, $D$ and $E$ 
can be determined from a complete phase
shift analysis of $NN$ data. In practice, for Glauber theory
calculations one chooses a parameterization of the central amplitude, $A$,
in terms of three experimentally known parameters: the total $NN$
cross section, the dif\/fraction slope $b_o^2$ and the ratio $\rho$
of the real to imaginary parts of the elastic forward scattering
amplitude. We discuss the differences between these approaches later
in Sec.~\ref{secnnparams}. For the following discussion, we employ
the latter approach, writing
\begin{equation}
A (l) = \frac{k \, \, \sigma_{tot}^{NN}}{4 \pi} (\rho + i)
\, \exp (-0.5 \, l^2 \, b_o^2) \,.
\end{equation}
Note that sometimes this amplitude is parameterized as 
\begin{equation}
A = A_o \cdot \exp(- \beta_A l^2)\, ,
\end{equation}
where $A_o$ and $\beta_A$ are complex
numbers, leading to a prof\/ile function of the form
\begin{equation}
\Gamma (\vec b) = \frac{ \sigma_{tot}^{NN} (1 - i \, \rho)}{4 \pi
b_o^2} \, \, \exp (- \frac{\vec b^2}{2 \, b_o^2}) \,.
\end{equation}
For a 1 GeV nucleon, $b_o \approx 0.5$ fm and accordingly the
prof\/ile function changes on a scale that is much smaller than the
nuclear radius, i.e., approximately $2$ fm for the deuteron. This
means that the central FSI operator is a short-ranged function with
respect to the transverse separation of the nucleons and a long-ranged
function with respect to their longitudinal separation.

In the analysis of the proton-nucleus scattering data, the spinless
version of Glauber theory, including only the central part of the $NN$
interaction, was very successful \cite{glaumat,wallace}. We also start
by including only the central part, as was done before in
\cite{mydeu,mytensor}.

\subsection{Central FSI}

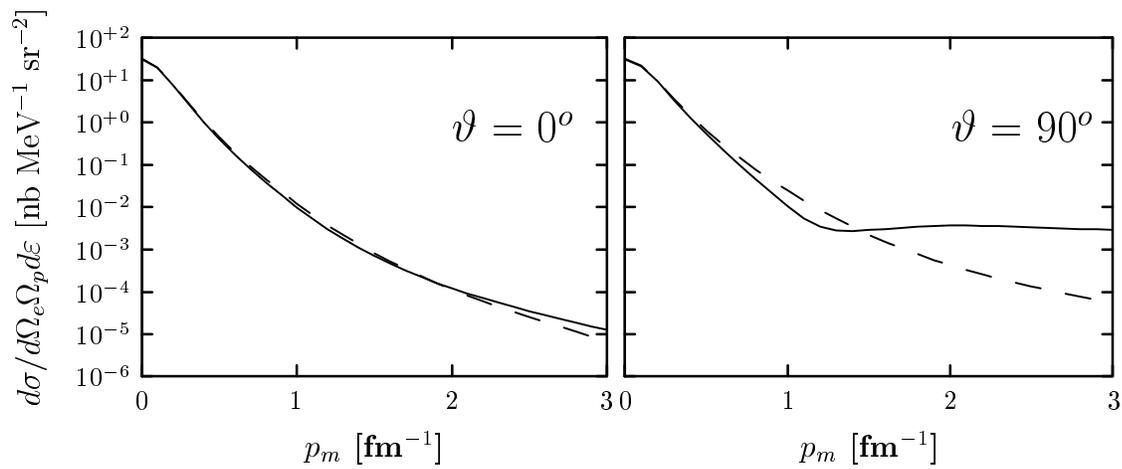
\begin{figure}
\caption{ On the left, the f\/ive-fold dif\/ferential cross section for the
reaction $D(e,e'p)n$ is shown in parallel kinematics (the angle of the
missing
momentum $\vartheta = 0^o$), on the right, it is shown for
perpendicular kinematics ($\vartheta = 90^o$). The solid line shows the
calculation including central FSI, while the dashed line shows the PWIA result.
The beam energy is $6$ GeV.
  \label{figwqfsipw}}
\end{figure}

Before we start the discussion of the results including FSI, some
remarks about conventions and the choice of kinematics are in
order. We plot our results versus the missing momentum $\vec p_m =
\vec q - \vec p_N$, where $\vec q$ is the momentum transferred to the
nucleus by the electron and $\vec p_N$ is the momentum of the detected
nucleon in the final state. In PWIA, it coincides with the negative
initial momentum of the struck nucleon inside the nucleus: $\vec p_m =
-\vec p$.  We denote the angle between $\vec p_m$ and $\vec q$ by
$\theta$, and the term ``parallel kinematics'' indicates $\theta =
0^o$ and ``perpendicular kinematics'' indicates $\theta = 90^o$.  In
this paper, we assume that the experimental conditions are such that
the kinetic energy of the outgoing nucleon and the angles of the
missing momentum, $\theta$ and the azimuthal angle $\phi$, are fixed.
The kinetic energy $T_p$ of the outgoing proton is fixed to 1 GeV. For
changing missing momentum, the transferred energy and momentum change
accordingly.  Fig.~\ref{figwqfsipw} $ $ shows the f\/ive-fold
dif\/ferential cross section for the reaction $D(e,e'p)n$ in parallel
kinematics (left panel) and for perpendicular kinematics (right
panel).  In perpendicular kinematics, the ef\/fect of the central FSI
is most pronounced: for missing momenta from $0.8$ fm$^{-1}$ to $1.5$
fm$^{-1}$, the FSI reduces the cross section, and for missing momenta
higher than $1.5$ fm$^{-1}$, it drastically enhances the cross
section, by up to an order of magnitude.  The central FSI ef\/fects in
parallel kinematics are smaller, but still important for higher
missing momenta.

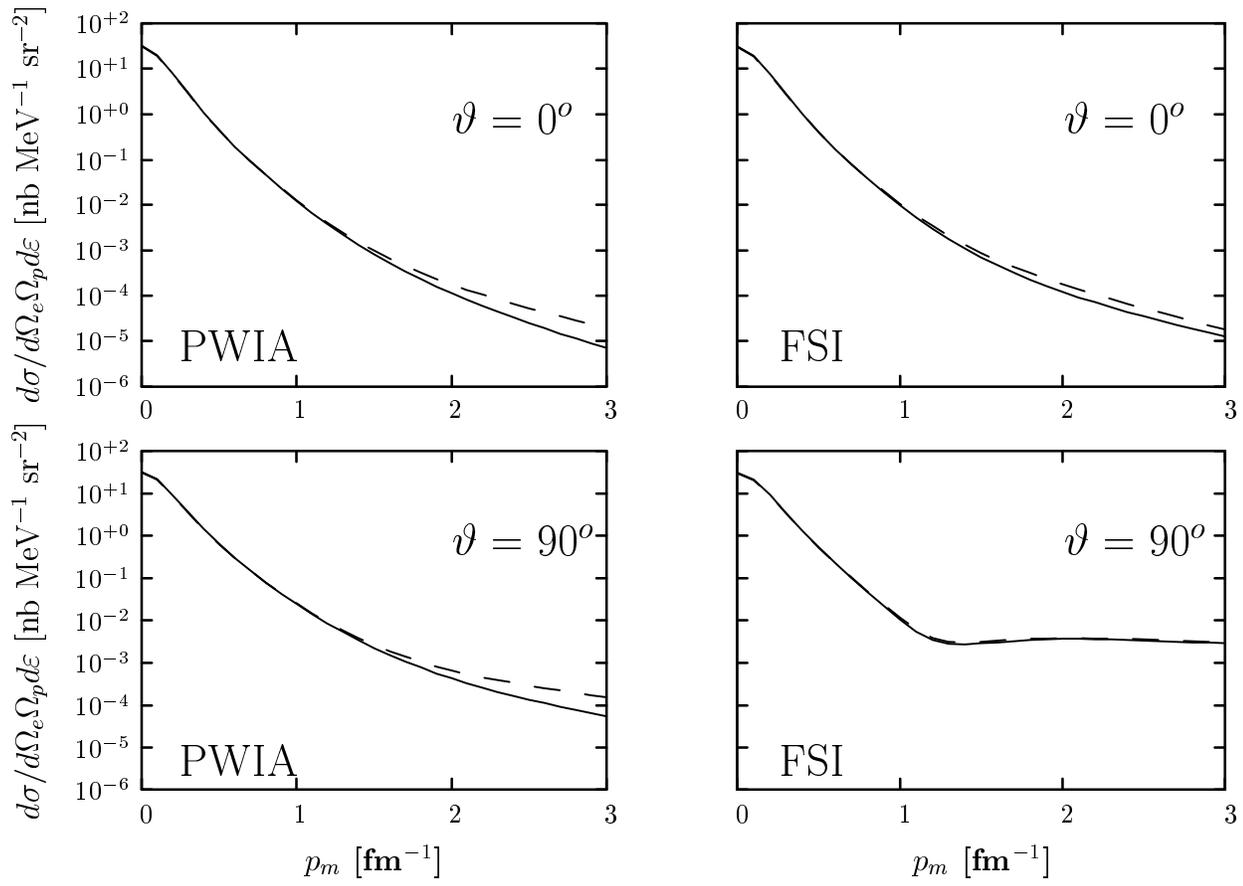
\begin{figure}
\caption{ On the left, the f\/ive-fold dif\/ferential cross section for the
reaction $D(e,e'p)n$ is shown in PWIA; on the right it is shown
including central FSI. The top panels show the cross section in
parallel kinematics, whereas the bottom panels show it in
perpendicular kinematics. The solid line shows the
calculation performed with the Bonn wave function, while the dashed line
is the Paris wave function result. The beam energy is $6$ GeV.
  \label{figwqmod}}
\end{figure}

Fig.~\ref{figwqmod} shows the cross section in PWIA and with central FSI,
calculated using the Bonn and the Paris wave functions. Whereas the
predictions of the two models dif\/fer for PWIA for higher missing
momenta, the curves become almost indistinguishable in perpendicular
kinematics and very similar in parallel kinematics once the FSI
ef\/fects are included. These results can be understood when we
consider the properties of f\/inal-state interactions: they are short-ranged, 
and this means that while they act on the S-wave, their ef\/fect
on the D-wave is small, as the D-wave is suppressed at short distances
by the centrifugal barrier.  In the discussion of Fig.~\ref{figmodnp},
we have seen that the dif\/ference between the models at higher
momenta stems from their dif\/ferent D-wave content, and that the
D-wave dominated the momentum distribution at higher momenta. Once FSI
are included in the calculation, they redistribute S-wave strength
from low and medium missing momenta to high missing momenta, so that
with FSI, the S-wave becomes important for this kinematic region. As
the wave function models do not dif\/fer too much in the S-wave, the
cross sections calculated with the two models including FSI ef\/fects
do not dif\/fer very much, either.  These issues have been discussed in
greater detail in \cite{mydeu}.  

Thus, the unpolarized $D(e,e'p)n$ cross section is not a good
candidate for the experimental discrimination between theoretical
models.  The central FSI ef\/fects in the longitudinal and transverse
response functions are fairly similar to the FSI ef\/fects in the
cross section (for the convenience of the reader, the definition of
the response functions and the different components of the current
operator is given in the appendix).  This comes as no surprise,
because $R_L$ is dominated by the zeroth-order charge operator and
$R_T$ is dominated by the magnetization current. Their coordinate
space structure is the same apart from dif\/ferent multiplicative
factors, the main dif\/ference being in the spin-operator
structure. As we are considering only central FSI at this moment, the
FSI ef\/fect in the two sectors will be the same.

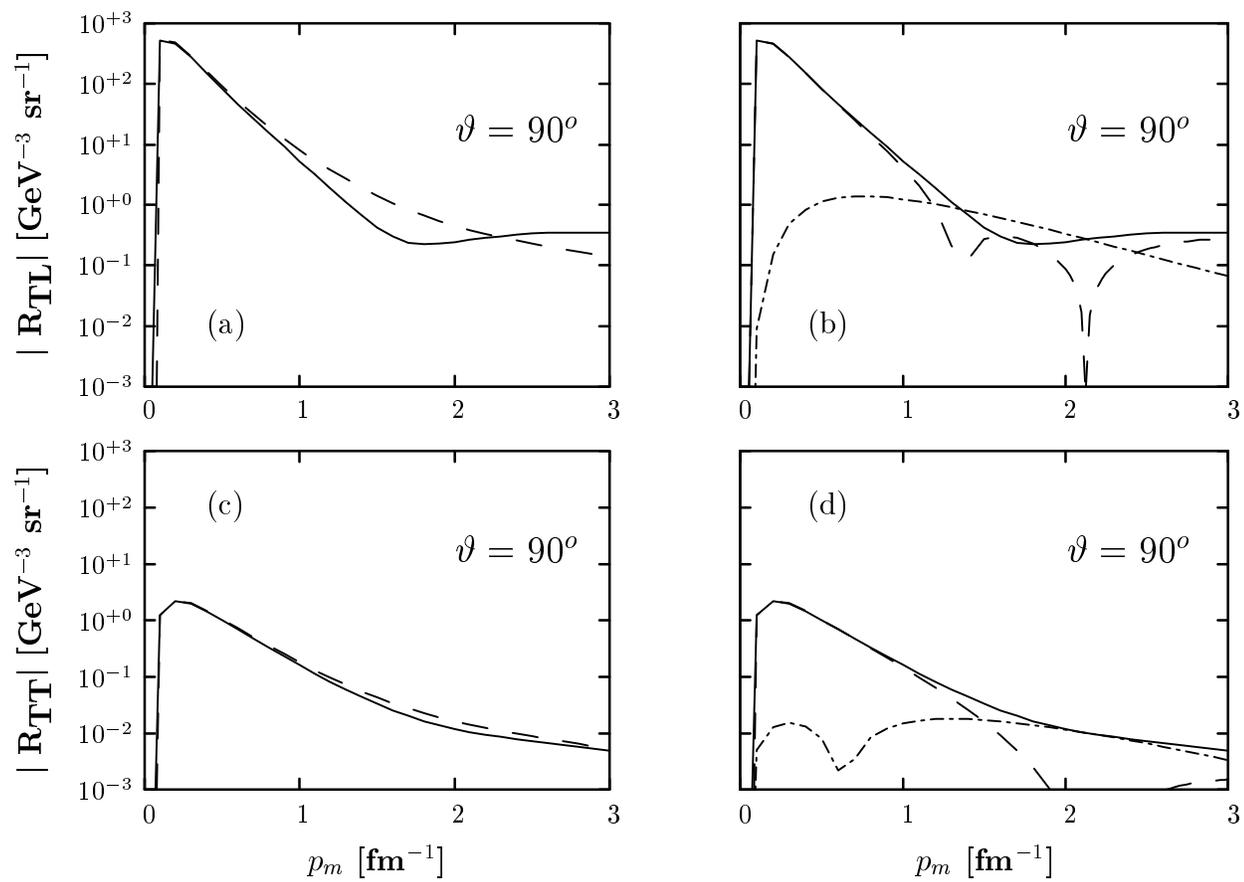
\begin{figure}
\caption{ The top panels, (a) and (b), show the absolute value of the
transverse-longitudinal response function $R_{TL}$, while the bottom panels,
(c) and (d), show the absolute value of the transverse-transverse
response function $R_{TT}$. In all panels, the solid line shows the
calculation including central FSI. The dashed line in (a) and (c) shows the
PWIA result, whereas the dashed line in (b) and (d) shows the S-wave
contribution and the dash-dotted line in (b) and (d) shows the D-wave
contribution. The total response $R_{TT}$ is negative over the entire
$p_m$ range, as is its S-wave part. The D-wave part is positive for
missing momenta less than $0.5$ fm$^{-1}$ and then turns negative, too.
The total response $R_{TL}$ is positive over the entire $p_m$ range,
as is its D-wave part. The S-wave part is negative from 
$1.4$ fm$^{-1}$ to $2.1$ fm$^{-1}$.
\label{figint1}}
\end{figure}

In order to find a better suited observable for the study of the
nuclear ground state, we turn our attention to the smaller
interference responses, $R_{TT}$ and $R_{TL}$, which have a dif\/ferent
structure than the larger responses $R_L$ and $R_T$. Since the former vanish in
(anti)parallel kinematics, in Fig.~\ref{figint1} we show them only in 
perpendicular kinematics. One should note that the transverse-longitudinal
response $R_{TL}$ is considerably smaller than $R_L$ and $R_T$, and
that in turn the transverse-transverse response $R_{TT}$ is much
smaller than $R_{TL}$. We remark that the separation of
the interference responses from the measured cross section is by no
means easy; here, we focus on discussions of their properties, disregarding the
experimental feasibility for the moment.

In panels (a) and (c) of Fig.~\ref{figint1} we show $R_{TT}$ and
$R_{TL}$ calculated in PWIA (dashed line) and including central FSI (solid
line); in panels (b) and (d), we show the responses calculated with
central FSI decomposed into their S-wave (dashed line) and D-wave (dash-dotted
line) components.  

First, we discuss $R_{TL}$, shown in panels (a) and (b) of Fig.~\ref{figint1}.
Starting at missing momenta of $1$ fm$^{-1}$, the
transverse-longitudinal response shows a strong FSI ef\/fect: at first,
the response is reduced by f\/inal-state interactions, then, for $p_m >
2.2 $ fm$^{-1}$, it is enhanced. The overall behavior of the FSI
ef\/fect is similar to that in the cross section and in $R_L$ and $R_T$,
but the enhancement due to FSI sets in at higher missing momentum ---
for the cross section, it starts at $p_m > 1.5$ fm$^{-1}$. The
decomposition of the response into S-wave and D-wave components, as shown
in panel (b), exhibits a unique feature of $R_{TL}$ in that even with the
f\/inal-state interactions included, the D-wave plays a prominent role
and at medium missing momenta, $1.2$ fm$^{-1} < p_m < 2.3$ fm$^{-1}$, 
it is much larger than the S-wave contribution. This does not
happen in the longitudinal response \cite{mydeu} or in the transverse
response.  The D-wave is scarcely af\/fected by FSI, and only when the 
S-wave contribution becomes larger than the D-wave contribution does the
typical enhancement of the response in perpendicular kinematics due to
FSI set in.

The transverse-transverse response $R_{TT}$ shows only a small ef\/fect
of FSI, namely, a slight reduction at medium missing
momenta. Fig.~\ref{figint1} (d) explains why: for missing momenta $p_m
> 1.5$ fm$^{-1}$, the response is entirely given by the D-wave
contribution, while the S-wave is negligible for medium missing momenta and
increases again only for $p_m > 3 $ fm$^{-1}$, a region where the
whole response is small. Since the central
FSI are short-ranged, they do not af\/fect the D-wave very much --- see the
discussion above --- and therefore the FSI ef\/fects on $R_{TT}$ are
small. 

So at this stage, it seems that, if they could be determined experimentally,
the interference responses might be well-suited for an
experimental discrimination between dif\/ferent theoretical models for
the nuclear ground state, as they dif\/fer mainly in their D-wave
content and both $R_{TT}$ and $R_{TL}$ show sensitivity to
it. However, here we have to be careful and review the
approximations which went into our calculation so far.
One has to keep in mind that the interference responses are much smaller
than the longitudinal response $R_L$ and the transverse response 
$R_T$, and that $R_{TT}$ roughly accounts for only $1 \%$ of the
total cross section, and $R_{TL}$ is not much bigger. So any
additional contribution that would be a $1 \%$ ef\/fect on the total cross
section, and which might therefore be neglected with good reason
in the calculation of the cross section, could give rise to a huge ef\/fect
in the interference responses, especially in $R_{TT}$. For example,
meson-exchange current effects (which are not considered in the present work)
may play such a role.

One of the reasons why the transverse-transverse response is so small
is that in PWIA the magnetization current cannot contribute, unless one
includes a second-order relativistic correction to the current, see
\cite{relcur}, with which it can interfere. Recalling that $R_{TT} = 2
\Re [ J^*_+ \, J_- ]$ and taking into account that for the
magnetization current, $J_{\pm}$ contains the spin-flip operator
$\sigma_{\pm}$, it is easy to see that for the S-wave, the
magnetization current cannot contribute, unless there is a
spin-dependent, spin-flipping FSI. This means
that, although it is quite reasonable and has been common
successful practice to use only a central f\/inal-state interaction
within the framework of Glauber theory in the calculation of cross
sections, it is not appropriate to neglect the spin-dependent FSI in
the calculation of certain responses, in particular in the calculation
of the interference responses discussed here. We now proceed to address
this issue.

\subsection{Spin-dependent FSI}

The $NN$ amplitude in Eq.~(\ref{eqnn}) contains one spin-orbit term
and three double-spin-flip terms.  The unique determination of these
contributions to the $NN$ amplitude is more difficult than for the
central part, and especially the double spin-flip terms are hard to
obtain. Due to their different spin structure the spin-orbit term and
the double spin-flip terms have different effects on the $(e,e'p)$
observables, so that a separate discussion of them is justified.
Here, we concentrate on the spin-orbit term. It may be parameterized
as
\begin{equation}
B (l) = \gamma \, \, \frac{k \, \, \sigma_{tot}^{NN}}{4 \pi} (\rho_s + i)
\, l \, \exp (-0.5 \, l^2 \, b_s^2) \,.
\end{equation}
The corresponding profile function reads:
\begin{equation}
\Gamma_s (\vec b) = - i \, \gamma \, 
\frac{\sigma_{tot}^{NN}} {4 \pi \, b_s^4}
\, (1 - i \rho_s) \, b \,  \exp (- \frac{b^2}{2 \, b_s^2}) \,,
\end{equation}
so that we have for the full profile function defined in 
Eq.~(\ref{defprof})
\begin{equation}
\Gamma (b) = \Gamma_c (\vec b) + \Gamma_s (\vec b) \sigma \cdot \nu \,, 
\end{equation}
with $\nu = \hat b \times \hat k$.  Comparing the expression for the
spin-orbit amplitude with that for the central amplitude, the main
dif\/ference is in the functional form of the former: it
contains a factor $l$, where $l$ is the transferred momentum, and this
makes the corresponding spin-orbit prof\/ile function $\Gamma_s (\vec
b)$ a longer-ranged function than the short-ranged central prof\/ile
function $\Gamma (\vec b)$.  In addition, the value of $b_s$ is larger
than that of $b_o$ used for the central amplitude, further
increasing the longer range of the spin-orbit amplitude.  The value
for $\gamma \approx 0.16$ fm taken from \cite{alkhazov} indicates
that the absolute value of the spin-orbit amplitude is much smaller than
the central contribution. This indicates that the
contribution of the spin-orbit f\/inal-state interaction will be
significant only if it interferes with the central FSI. 

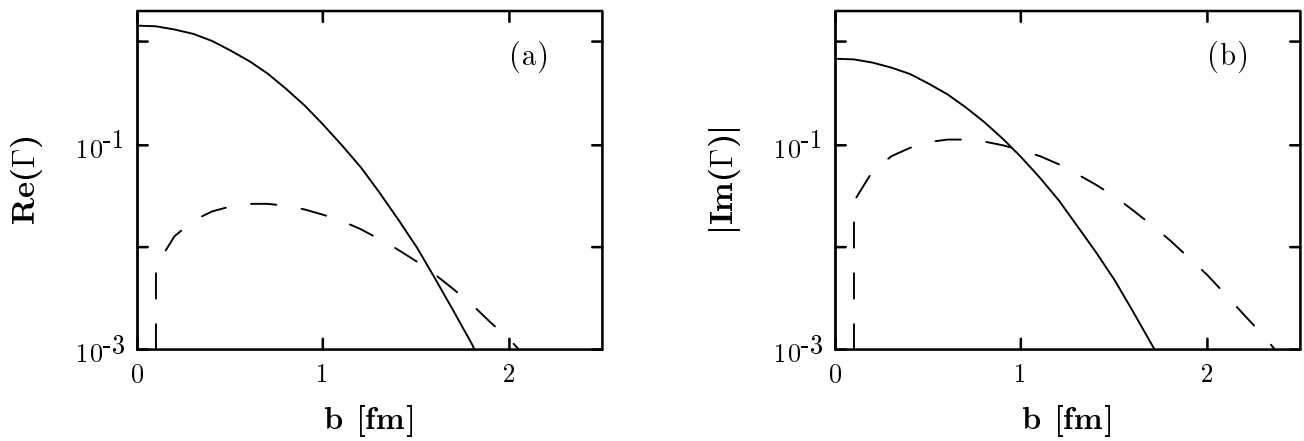
\begin{figure}
\caption{ The real part (left panel) and the absolute value of the
imaginary part (right panel) of the central profile function (solid
line) in coordinate space for $pn$ scattering at $T_{lab} = 1$ GeV and
the spin-orbit profile function (dashed line). The profile functions
are plotted versus the transverse separation $b$ of the two nucleons. The
parameters for the central part are $\sigma_{tot}^{pn} = 41.1$ mb,
$b_o = 0.48$ fm, and $\rho = -0.48$. For the spin-orbit part, we use
$\gamma = 0.16$ fm, $b_s = 0.65$ fm and $\rho_s = -0.24$.
The imaginary part of the spin-orbit profile function is negative and
the imaginary part of the central profile function is positive.}
\label{figprofile}
\end{figure}

To illustrate these points, in Fig.~\ref{figprofile} we compare the
real and imaginary parts of the central profile function (solid line)
and the spin-orbit profile function (dashed line) for $p-n$ scattering.
We use the parameters $\sigma_{tot}^{pn} = 41.1$ mb, $b_o = 0.48$ fm
and $\rho = -0.48$, while for the spin-orbit part we use $\gamma = 0.16$
fm, $b_s = 0.65$ fm and $\rho_s = -0.24$. These parameters are the
average of the parameters quoted in Table 10 of Ref. \cite{wallace}
for Glauber theory analyses at $T_{lab} = 1$ GeV. For the real parts
of the profile functions, shown in panel (a), the central contribution
is dominant for transverse separations $b < 1.6$ fm, then falls off
quite fast and at $b \geq 2.5$ fm is two orders of magnitude
smaller than the spin-orbit contribution. However, at these larger
separations, the most important contribution stems from the imaginary
part of the spin-orbit profile function; see panel (b). It dominates
the central contribution already for $b > 1$ fm, and remains quite
large for larger $b$.  We note that there are even
larger values for $b_s$ published in the literature, e.g. a value of
$b_s = 0.81$ fm in \cite{alkhazov}, and that, if one substitutes one
of these for our
smaller value of $b_s = 0.65$ fm, the spin-orbit profile function
will be enhanced significantly at larger $b$. 

In the following, we investigate the consequences of the longer-range 
character and different spin structure of the spin-orbit
profile function.  For the cross section and the two large responses,
$R_L$ and $R_T$, the influence of the spin-orbit FSI is --- as expected
--- very small.  The simple structure of the longitudinal response, $R_L
= |\rho|^2$, and transverse response, $R_T = |J_+|^2 + |J_-|^2$,
does not restrict any kind of contribution even with central FSI,
and therefore, by including the spin-orbit FSI which would allow for
different spin structures to contribute, no strength is gained.  The
results with and without spin-dependent FSI
almost coincide, and therefore we do not show them here.

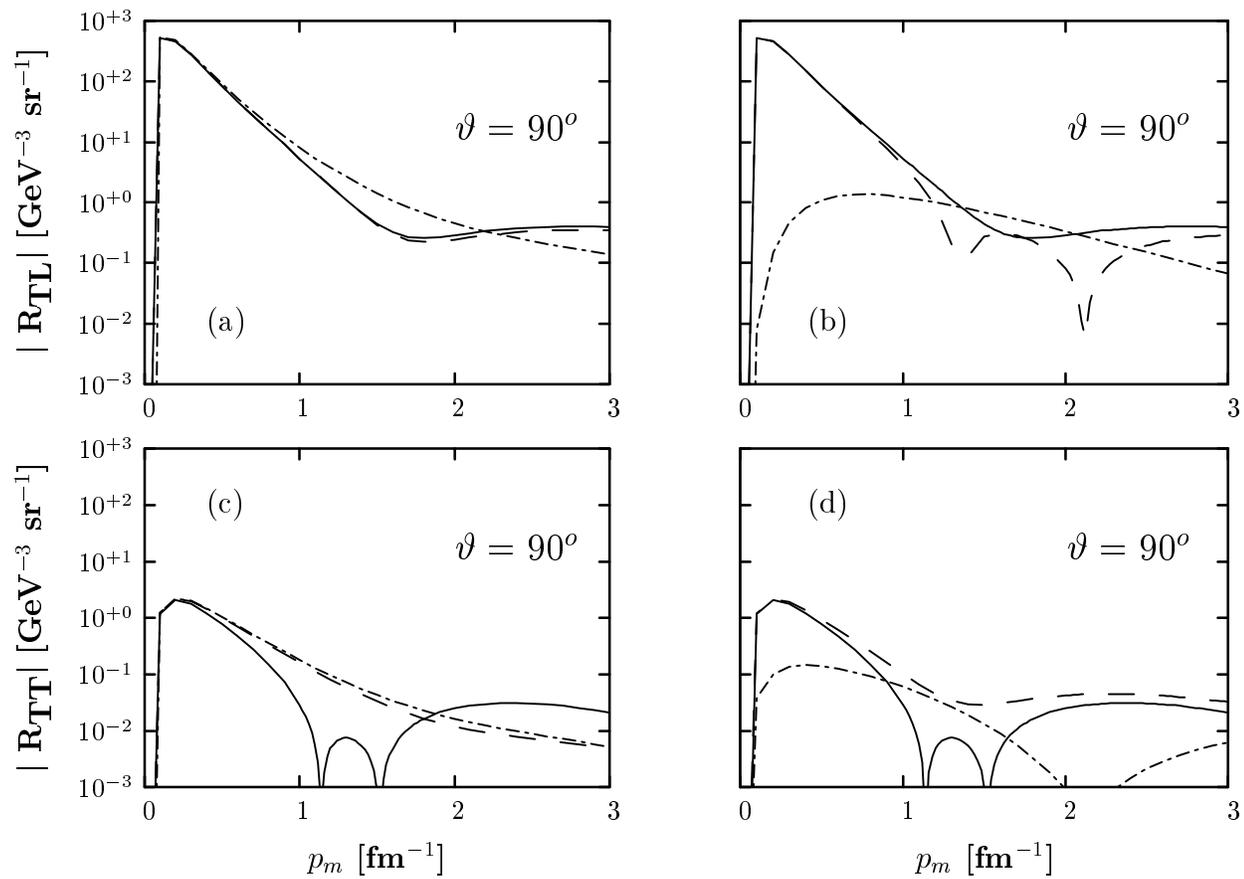
\begin{figure}
\caption{ The top panels, (a) and (b), show the absolute value of the
transverse-longitudinal response function $R_{TL}$, while the bottom panels,
(c) and (d), show the absolute value of the transverse-transverse
response function $R_{TT}$. In all panels, the solid line shows the
calculation including central and spin-orbit FSI. The dashed line in
(a) and (c) shows the central FSI result, and the dash-dotted line in
(a) and (c) shows the PWIA result.  In panels (b) and (d), the dashed
line shows the S-wave contribution and the dash-dotted line shows the
D-wave contribution. The full response $R_{TT}$ is negative, except
for $1.2$ fm$^{-1} < p_m < 1.5$ fm$^{-1}$. Its S-wave part is negative
over the entire $p_m$ range, and its D-wave part is positive over the
entire $p_m$ range . The full response $R_{TL}$ is positive over the
entire $p_m$ range, as is its D-wave part. The S-wave part is negative
from $1.4$ fm$^{-1}$ to $2.1$ fm$^{-1}$.
\label{figint2}}
\end{figure}

The situation changes once we consider the interference responses.
They are shown in Fig.~\ref{figint2}. On the left-hand side, the
responses $R_{TT}$ and $R_{TL}$ are shown calculated with the full FSI
(solid line), with central FSI only (dashed line), and without FSI
(dash-dotted line). For the transverse-longitudinal response, the
additional spin-orbit FSI has only a marginal effect on the result for
missing momenta from $1.8$ fm$^{-1}$ to $3.5$ fm$^{-1}$.
Therefore, also the decomposition into S-wave and D-wave contributions
is practically unchanged compared to the central FSI results.

For the transverse-transverse response, the spin-orbit FSI plays an
important role. The results for PWIA, i.e. no FSI at all, and for
central FSI are very similar; they show a smooth, structureless
decrease with increasing missing momentum, and the response is
negative for the entire $p_m$ range. Once the spin-orbit FSI is
included, the response starts out negative at small missing momenta,
then changes sign at $p_m = 1.2$ fm$^{-1}$, and becomes negative again
at $p_m = 1.5$ fm$^{-1}$.  For higher missing momenta $p_m > 2$
fm$^{-1}$, the spin-orbit FSI increases the response significantly.
This is due to the fact that with the spin-dependent FSI, the
magnetization current now can also have an S-wave contribution to
$R_{TT}$.  Without spin-dependent FSI, this contribution does not
exist. As the magnetization current is the largest contribution to
$J_+$ and $J_-$, and as the S-wave is the largest component of the
deuteron wave function, the new contribution introduced by the
spin-orbit FSI is obviously important. This is demonstrated in
Fig.~\ref{figint2} (d), where the full transverse-transverse response
is decomposed into S-wave (dashed line) and D-wave
(dash-dotted line) contributions. The S-wave contribution is larger
than the D-wave contribution for all missing momenta, and it is
dominant for $p_m > 2$ fm$^{-1}$. This is in sharp contrast to the
picture for central FSI only (compare with Fig.~\ref{figint1} (d)):
there, the S-wave contribution is much smaller than the D-wave for
$p_m >2$ fm$^{-1}$, while it is practically negligible in the region from 2
fm$^{-1}$ to 2.5 fm$^{-1}$. Indeed, the increase in the S-wave part in this
region amounts to more than 1.5 orders of magnitude. The effect of the
spin-orbit FSI on the S-wave part can be attributed to the spin
structure changing nature of the FSI. 

There is another interesting
spin-orbit FSI effect here which originates in its longer range, namely, the
effect on the D-wave. As mentioned above and discussed in detail in
\cite{mydeu}, the central FSI does not affect the D-wave very much, as
it is short-ranged and the D-wave is suppressed at short distances by
the centrifugal barrier. However, the spin-orbit FSI has a longer
range, and if we compare the D-wave part of $R_{TT}$ with central FSI
(Fig.~\ref{figint1} (d)) and with central + spin-orbit FSI (Fig.~\ref{figint2}
(d)), there is a dramatic difference: the latter causes the 
D-wave part to be positive over the entire $p_m$ range, whereas with
central FSI only, the D-wave part is positive for missing momenta less
than $0.5$ fm$^{-1}$ and turns negative afterwards. The absolute
magnitude is affected as well by the spin-orbit FSI: it introduces a dip
around $p_m = 2.2$ fm$^{-1}$, whereas with the central FSI it just
decreases smoothly.  This is the first instance we encounter where the
FSI changes the D-wave contribution significantly.  In combination
with the effect of the spin-orbit FSI on the S-wave, this leads to an
inversion of the importance of the two contributions for
$R_{TT}$. This is very important, as the ability to discriminate
experimentally between different theoretical models for the nuclear
ground state depends on being sensitive to the D-wave, in which the
model predictions differ.

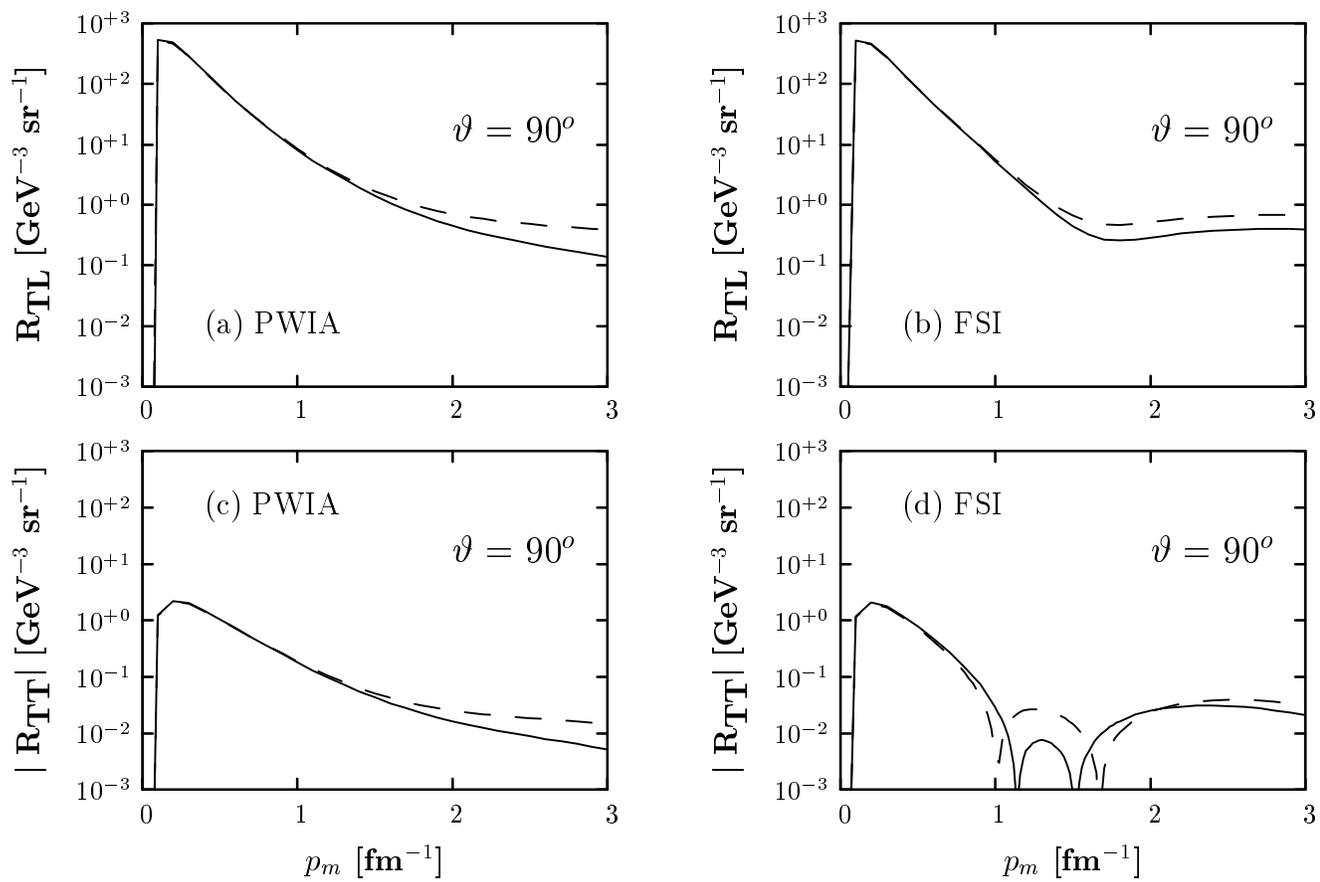
\begin{figure}
\caption{ On the left, the interference responses $R_{TL}$ and $R_{TT}$
are shown in PWIA; on the right they are shown including central and
spin-orbit FSI, in both cases for perpendicular kinematics. The top
panels show the transverse-longitudinal response function, and the bottom
panels show the transverse-transverse response function. The solid
line shows the calculation performed with the Bonn wave function, whereas the
dashed line is the Paris wave function result.
\label{figint2mod}}
\end{figure}

This leads to the last point in our
discussion of the interference responses: the comparison of the
different model wave functions.  In Fig.~\ref{figint2mod}, in panels
(a) and (c), we show the transverse-longitudinal and 
transverse-transverse responses in PWIA, calculated with the Bonn wave
function (solid line) and the Paris wave function (dashed line).  As
expected, the results differ at higher missing momentum where we have
seen that the D-wave plays an important role. Once the full FSI,
including central and spin-orbit interactions, is included in the
calculation, $R_{TL}$ is affected at higher missing momenta; however, it
basically retains its potential for discrimination between the
different ground-state wave functions. This is due to the fact that
for $R_{TL}$, the D-wave is an important contribution for $p_m > 1.3$
fm$^{-1}$. For the transverse-transverse response, the difference in
the PWIA results is roughly the same as for $R_{TL}$.  However, the
full FSI alters the picture: for higher missing momenta, $p_m > 2$
fm$^{-1}$, the Bonn and Paris wave functions give very similar
results. At medium missing momenta, $1$ fm$^{-1} < p_m < 1.8$ fm$^{-1}$,
the sensitivity to the models is increased by FSI. This is the region
where S-wave and D-wave contributions have similar size. 
Unfortunately, the practical use of this information is limited: 
from the experimental point of view, a separation of $R_{TT}$ is very
dificult,
as this is the smallest of the four responses generally present in the
unpolarized $D(e,e'p)n$ reaction, and the experimental errors might
be largest in the region of  $1$ fm$^{-1} < p_m < 1.8$ fm$^{-1}$, which
includes two zeros of the response. From the theoretical point of view,
it is advisable to include the full relativistic electromagnetic current
for the calculation of $R_{TT}$, see \cite{relcur}, and it is possible
that meson-exchange currents may play a role, too. In this paper,
it is our goal to demonstrate that even in the GeV energy regime, it
is mandatory to include spin-dependent FSI effects, although we do not claim
to have treated the reaction mechanism fully. 

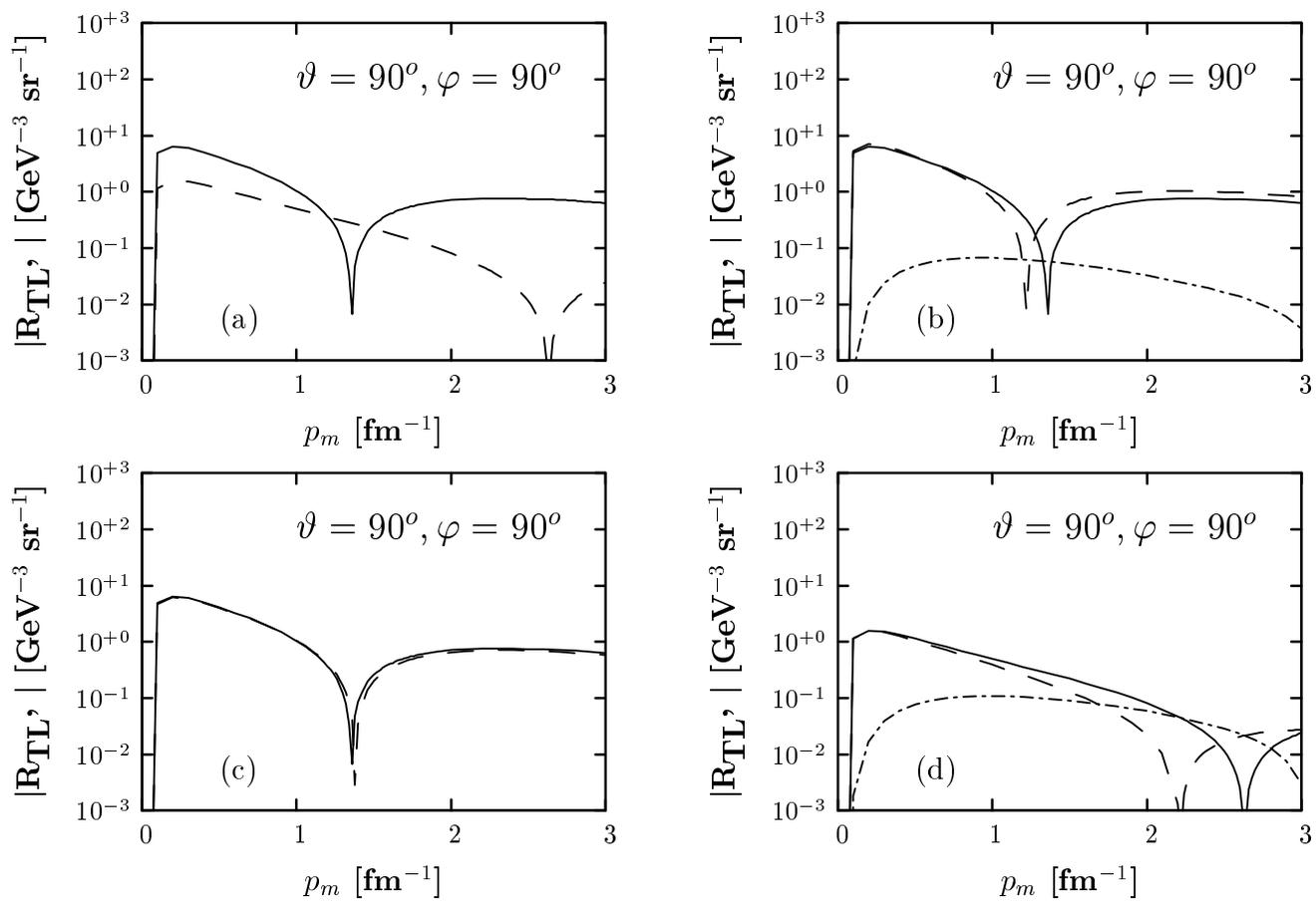
\begin{figure}
\caption{The fifth response function, $R_{TL'}$, is shown in panel (a)
calculated with the full FSI (solid line) and with central FSI only
(dashed line) in perpendicular kinematics at an azimuthal angle
$\varphi = 90^o$.  The full response is negative for $p_m < 1.4$
fm$^{-1}$; with central FSI only, it is positive up to $p_m = 2.6$
fm$^{-1}$. Panel (c) shows the response calculated including full FSI
with the Bonn wave function (solid line) and the Paris wave function
(dashed line). In panels (b) and (d), besides the full response (solid
line), we show the S-wave part (dashed line) and the D-wave part
(dash-dotted) for the calculation with central FSI (d) and full FSI
(b).}
\label{figfifth}
\end{figure}

We now turn to the fifth response function, $R_{TL'}$, which can only be
measured with a polarized electron beam and vanishes for PWIA. Its
dependence on the out-of-plane angle $\varphi$ is given by $\sin
\varphi$, and therefore we consider this response at an out-of-plane angle
$\varphi = 90^o$.  Fig.~\ref{figfifth} (a) shows $R_{TL'}$ calculated
with the full FSI (solid line) and with the central FSI (dashed line).
With central FSI, the response is positive up to $p_m = 2.6$ fm$^{-1}$
and then changes sign. Once the spin-orbit FSI is included, the
response is negative for $p_m < 1.4$ fm$^{-1}$ and becomes positive
afterwards.  Also, the magnitude is changed drastically by the 
spin-dependent FSI: for small missing momenta, $R_{TL'}$ is increased by a
factor of four and more. For the higher missing momenta, the response
is increased by more than one order of magnitude. Being rather large, 
this is likely to be an observable effect, and it shows clearly that for
polarization observables such as the fifth response function the inclusion
of spin-dependent FSI is absolutely necessary.

In Figs.~\ref{figfifth} (b) and (d), we show the decomposition of the full wave
function into S- and D-wave contributions, for the full FSI in panel
(b) and for central FSI only in panel (d). In both cases, the S-wave
is the most prominent contribution, especially for the full FSI
calculation, where the influence of the D-wave is reduced to the
region around $p_m \approx 1.3$ fm$^{-1}$, i.e., where the S-wave
contribution changes sign. In the central FSI calculation, the D-wave
contribution is slightly more important, as its size is larger than
for the full FSI result and as the absolute magnitude of the S-wave
contribution is smaller than for the full FSI at higher missing
momenta. Consequently, the difference between the Bonn and Paris wave
function results is marginal; see Fig.~\ref{figfifth} (c), where the
full line represents the Bonn result and the dashed line represents
the Paris result. The two curves are almost on top of each other.

The large differences between the full FSI and central FSI results
stem from the different spin-structure of the spin-orbit FSI operator.
For central FSI, there are two contributions in the S-wave: the
largest is given by the magnetization current and the spin-orbit
charge operator, and the convection current and the zeroth-order
charge operator yield a contribution of different sign and
approximately half the magnitude of the magnetization/spin-orbit
contribution (for an extensive discussion of the electromagnetic
current see \cite{relcur}).  Once the spin-orbit FSI is included, the
previously existing contributions are unchanged, but there arise new
contributions, the dominant new contribution being the one coming from
the magnetization current and the zeroth-order charge operator. It is
about $40 \%$ larger than the magnetization current/spin-orbit charge
operator result, and it has a different sign. This new contribution is
responsible for changing the sign and increasing the magnitude of the
S-wave contribution to the fifth response. There are also new contributions
involving the interference of the convection current with the spin-orbit
charge operator and the first-order convective spin-orbit current which
did not contribute at all for the central FSI, although these contributions
are small. The main effect is that the two biggest components of the
current, the zeroth-order charge operator and the magnetization current,
can interfere with each other and contribute due to the spin-orbit FSI,
and therefore change the character of the fifth response.

Previous calculations employing Glauber theory for $(e,e'p)$ reactions
at high energies only took into account central FSI
\cite{mydeu,mytensor,fs,kys}, which is reasonable when one is
interested in the cross sections, large responses and nuclear
transparencies.  However, the measurement of the smaller responses and
of the polarization observables is expected to reveal most about the
nuclear wave function, and so the inclusion of spin-orbit FSI will be
necessary for the interpretation of the experimental data.  There is
also a recent publication \cite{iks} in which both central and
spin-orbit FSI are treated within a Dirac eikonal formalism.  Although
spin-orbit FSI in \cite{iks} is also found to be important for certain
polarization observables and in $R_{TT}$, its effects in the fifth
response are small. This may be due to the consideration of a
different target nucleus, $^{16}O$, or due to the different approach
used there.

\section{Influence of the NN parametrizations}
\label{secnnparams}
In this section, we discuss the different parametrizations for
the $NN$ amplitude and the effects of changes in these parameters on
the results we have obtained. For the central part of the amplitude,
this has been done before in \cite{thesis}. There, it was shown that
changes in the slope parameter $b_o$ affect the results in
perpendicular kinematics, increasing them for high missing momenta if
$b_o$ is smaller and the interaction therefore shorter-ranged, and
decreasing them at high $p_m$ if $b_o$ is chosen larger. In any case,
the results with FSI lead to a considerably larger high missing
momentum tail than in PWIA. A change in $\rho$, the ratio of the real
to imaginary parts of the forward elastic scattering amplitude,
influences the results only in parallel and antiparallel kinematics.

The spin-dependent part of the amplitude is not known as well as the
central part, and there are many different values for the spin-orbit
parametrization to be found in the literature (for an overview, see
\cite{wallace}). The two main groups are the values derived from phase
shift analyses and from proton-nucleus Glauber theory calculations.
In the Glauber approach, one usually neglects all double-spin-flip
contributions and in addition assumes that the slope parameters are
real, in contrast to the complex values derived in the phase shift
analyses. Actually, the two different approaches already differ quite
significantly in the central part: the diffraction slope
derived in a typical phase shift analysis is about 50\% larger than the
one derived from Glauber calculations.  Characteristic values at $T_{lab} =
1$ GeV are listed in Tables \ref{tabnng} and \ref{tabnnp}. The phase shift
values are taken from \cite{mrw}, while the Glauber values are the average
of the parameters quoted in Table 10 of Ref. \cite{wallace}.  Except
for $\rho_s$, which enters in the spin-orbit part, all of the parameters
agree within $10 - 20 \%$. As already mentioned, the parameters $b_o$
and $b_s$ of the phase shift analysis have an imaginary part, although it
is small. For $\rho_s$, there is a wide spread of values even among the
different Glauber analyses, yielding values from $0.4$ to $-0.62$.
Clearly, we need to investigate how much a change in this
parameter affects our results.

The situation is somewhat more complicated by the fact that the 
final-state interaction is not calculated in the $N N$ c.m. 
system, for which the parametrizations are given, but in a system
where the incoming and outgoing nucleons all have different momenta:
before the FSI takes place, the proton carries the
sum of its initial momentum and the momentum $\vec q$ transferred by
the photon, and the neutron still carries the momentum it had
initially in the deuteron. After the FSI took place, the
neutron carries the missing momentum $\vec p_m$, and the proton has
the momentum $\vec {p_N}$. As the proton's momentum is high in any case,
and the neutron's initial momentum on the average will be smaller than
the Fermi momentum ($p_F \approx 55$ MeV for the deuteron), the
situation is rather asymmetric. At least for the deuteron, the
rescattering takes place in a system quite close to the lab system, as
the Fermi momentum is so small.

Therefore, we have to boost the $N N$ parameters from the c.m. system
to the appropriate ``rescattering system''. As in the rescattering
system the momenta of the nucleons do not transform into one another
under time reversal, we pick up different coefficients $C_1$ and $C_2$
for the spin-orbit part for the two different nucleons. This is in
analogy to the transformation from the c.m. frame to the Breit frame,
which was discussed in \cite{wallace,mrw}. We have transformed the $N
N$ parameters to the lab frame, since taking into account the nonvanishing
initial momentum of the neutron would only lead to negligible
additional corrections.  We used the method outlined in \cite{mrw} for
the boost.  It is well known that the central parameters do not change
in a transformation from the c.m. to the lab system, provided they are
multiplied with the properly boosted quantities in each reference
frame \cite{franglau}. This leaves the spin-orbit parameters (and also
the double-spin-flip terms, which we do not consider here). The
spin-orbit parameters now differ for the two nucleons, and the slope
parameters pick up an imaginary part. However, we carried out a
calculation with the imaginary parts of $b_s$ set to zero and found
that the results practically coincide with the results obtained for
the nonvanishing imaginary parts. The spin-orbit values for the lab
system obtained from the Glauber $p-A$ analysis are shown in Table
~\ref{tablab}. It is obvious that the boost to the lab system has a
very small effect on $\gamma$ and $b_s$.  The boost has a significant
effect only on $\rho_s$, which is almost zero for the neutron and
takes on about twice its c.m. frame value for the proton. The
spin-orbit values for the lab system obtained from the phase shift
analysis are shown in Table ~\ref{tablabpsa}.  Again, the boost
effects are significant only for $\rho_s$.

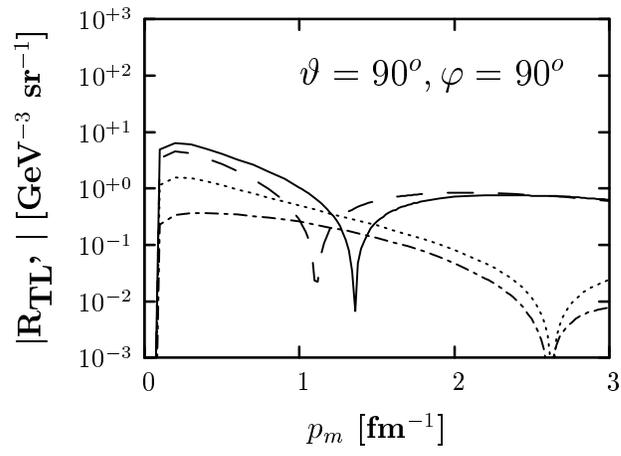
\begin{figure}
\caption{The fifth response function, $R_{TL'}$, is shown in
perpendicular kinematics at an azimuthal angle $\varphi = 90^o$.  The
various curves represent a calculation with the full FSI (solid line),
a calculation with central FSI and spin-orbit FSI only on the proton
(dashed line), a calculation with central FSI and spin-orbit FSI 
only on the neutron (dash-dotted line), and a calculation with
central FSI only (dotted line).}
\label{figpnsofsi}
\end{figure}

Instead of discussing the merits of the Glauber $p-A$ analysis versus
the phase shift analysis parameters, we investigate the sensitivity of
the final results to changes in the $N N$ parameters. From the
preceding paragraph, it is clear that we have to consider moderate
changes in $\gamma$ and $b_s$, and large changes in the less well
known $\rho_s$. As $\gamma$ simply multiplies the spin-orbit term, it
is obvious that a slight increase or decrease in $\gamma$ leads to a
slight increase or decrease in the overall spin-orbit FSI effect,
independent of the specific kinematics. We therefore focus on $b_s$
and $\rho_s$. As the fifth response is most sensitive to the
spin-orbit FSI, we choose it as a testing ground. Before checking the
sensitivity to the $N N$ parameters, it is useful to consider the
spin-orbit FSI effect on the proton and the neutron separately.  In
Fig.~\ref{figpnsofsi}, the solid line shows the fifth response
calculated with the full FSI, i.e. with the spin-orbit FSI acting both
on proton and neutron. The dashed line shows the results with the
spin-orbit FSI acting on the proton only. One can see that the
qualitative agreement between the curves is good. Naturally, the
influence of the spin-orbit FSI on the neutron is largest when the
fifth response changes sign, but for small missing momenta $p_m < 0.5$
fm$^{-1}$ and the high missing momentum tail, $p_m > 1.8$ fm$^{-1}$,
its role is minor. The situation changes when we switch off the
spin-orbit FSI on the proton, and keep only the spin-orbit FSI on the
neutron, as depicted by the dash-dotted line: the results are
qualitatively similar to the results with central FSI only (dotted
line) --- indeed, they are even somewhat smaller.

Note that here the case is different from proton-nucleus scattering,
where the spin-orbit FSI on the spectator nucleon does not contribute
for spin-0 targets. Here, it does contribute, although its
contribution is much smaller than that of the spin-orbit FSI on the
proton. This asymmetry is introduced in the problem by the assumption
of the impulse approximation, i.e. the assumption that the nucleon
with which the virtual photon interacted is the one which is detected
later. This means that the electromagnetic current operator acts only
on the proton, and therefore the spin-effects on proton and neutron
cannot be the same, as the electromagnetic current operator contains
spin-flipping parts.  For the conditions we assume here, namely a very
energetic virtual photon and a very energetic proton ($T_p = 1 $ GeV)
which is measured, the impulse approximation is a realistic
assumption, because it is quite unlikely that after its 
interaction with the photon the nucleon will transfer the momentum and energy
gained in the hard vertex completely to the other nucleon. For lower
energies, where both the proton and neutron in the final state have
similar momenta, the Born graph does contribute and the spin-orbit FSI
on the neutron should gain importance, as the situation then is more
symmetric for the two nucleons.

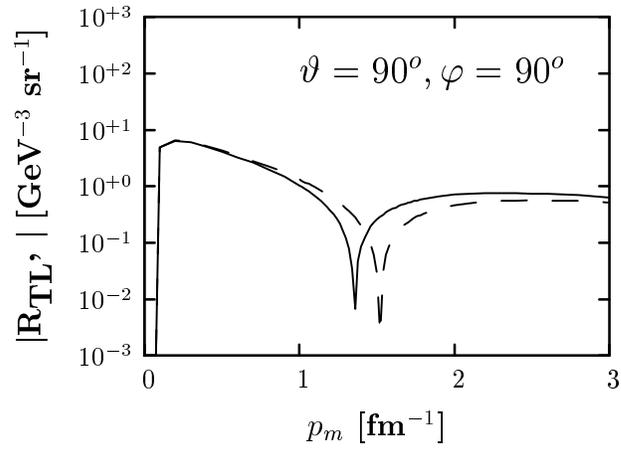
\begin{figure}
\caption{The fifth response function, $R_{TL'}$, is shown in
perpendicular kinematics at an azimuthal angle $\varphi = 90^o$.  The
solid line shows the results of the calculation with $\rho_s = 
-0.44$ for the proton, while the dashed line shows the result for
the changed sign of $\rho_s$, i.e. for $\rho_s = +0.44$.}
\label{figrhoso}
\end{figure}

In view of the importance of the spin-orbit FSI on the proton and the
relatively small role of the spin-orbit FSI on the neutron, we will only
discuss the sensitivity to the different $N N$ parameters for the
spin-orbit FSI on the proton. As the effects there are large,
this should allow for the easy identification of the effects of
changes in $b_s$ and $\rho_s$.  We start with different values of
$\rho_s$ for the proton, keeping all of the other $N N$ parameters the
same. The results of a calculation with $\rho_s = -0.44$ (the value we
used for all of the above calculations) are compared to the results of a
calculation with $\rho_s = + 0.44$ in Fig.~\ref{figrhoso}.  The
results are very similar to each other, the only significant
deviations occurring in the region where $R_{TL'}$ changes its sign.  We
conclude that even the very drastic change of $200 \%$ in the value of
$\rho_s$ does not alter the results qualitatively and even does not
alter the results quantitatively, except for the region of $1.3 $
fm$^{-1} < p_m < 2$ fm$^{-1}$, where the response has a zero and any
slight change in the calculation must induce a visible difference.
So, although the value of $\rho_s$ is not known with much
precision, this uncertainty does not influence the practical
calculations in a significant way.

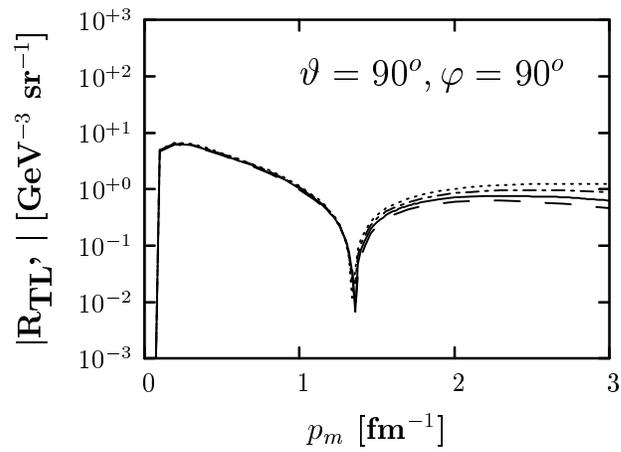
\begin{figure}
\caption{The fifth response function, $R_{TL'}$, is shown in
perpendicular kinematics at an azimuthal angle $\varphi = 90^o$.  The
solid line shows the results of the calculation with $\Re(b_s) = 0.65$
fm for the proton, the dashed line shows the result for $\Re(b_s) =
0.71$ fm, the dash-dotted line corresponds to $\Re(b_s) = 0.59$, and
the dotted line represents $\Re(b_s) = 0.52$.}
\label{figbs}
\end{figure}

In Fig.~\ref{figbs} we investigate the influence of the slope
parameter value on the fifth response. The solid line shows the result
with $\Re (b_s) = 0.65$ fm, while the dashed line represents the results for
$\Re(b_s) = 0.71$ fm, which makes the interaction even longer-ranged. 
As expected, and as observed for the central FSI and the slope
parameter $b_o$ in \cite{thesis}, the longer-ranged interaction
decreases the result at higher missing momenta. Consequently, the
shorter-ranged values of $b_s$ lead to a slightly increased result at
higher missing momenta, namely the dash-dotted curve for $ \Re (b_s) =
0.59$ fm and the dotted curve for $\Re (b_s) = 0.52$ fm. On an
absolute scale, these changes are small and they are present only for
$p_m > 1.5$ fm$^{-1}$. 

From these observations on the sensitivity to the $N N$ parameters, we
can conclude that the overall effect of the boost from
the c.m. frame of the $N N$ system to the lab frame only has
a small effect, quite similar to the observations made in \cite{mrw}
for proton-nucleus scattering and the transformation from the c.m. to
the Breit system. Also, the change in the parameters due to the boost
is smaller than the uncertainty in the value of the parameter.
Although the differences introduced by changing one $N N$ parameter at a  
time are small, we found that a simultaneous change in the values of all    
$N N$ parameters can add up to a significant shift of the zero of the
fifth response.

In this paper, we did not consider double-spin-flip terms in the FSI.
They are smaller than the spin-orbit term, and in the proton-nucleus
Glauber calculations they are usually neglected. We estimate that
their influence on the observables calculated here will be very small,
since the spin-orbit FSI already allow the ``big'' components of the
current and the wave function to contribute. Specifically, 
in the fifth response the contribution of the interference
between magnetization current and zeroth-order charge operator is the
biggest contribution one can expect, and it is already present when
the spin-orbit FSI is included. However, there may be other
observables, e.g. double polarization observables, where the 
double-spin-flip FSI might contribute --- in these cases, it will be necessary
to re-examine many of the approximations made in the present study.

\section{Summary and Outlook}

The f\/inal-state interactions have a rich structure: due to the
short-ranged nature of the central part of the FSI, they tend to smear
out the dif\/ferences between the various theoretical models for the
nuclear ground state. Our results indicate that one of the smaller
responses or a polarization observable might be better suited for this
purpose. A reliable theoretical calculation of these responses must
include the spin-orbit FSI as well as the central FSI.  We have
included the spin-orbit FSI, for the fifth response function showing
that the spin-dependent FSI is indeed crucial. The calculations
presented here go beyond the approach usually taken in calculations
for $(e,e'p)$ reactions in the framework of Glauber theory, where only
the central FSI is included.
In addition, we have pointed out that due to the
longer-ranged nature of the spin-orbit final-state interaction, the
FSI can also have a significant impact on the D-wave contribution.
The experimental measurement of the small responses and the
polarization observables is a challenge that holds potential to shed
light on interesting issues involving all of the ingredients in
electronuclear physics, namely, initial- and final-state nuclear
structure and electromagnetic operators.  Of specific relevance for
the present work, we note that the fifth response function will be
measured for a variety of nuclei at Jefferson Lab in the near future
\cite{hersman}.  

\acknowledgments 

S. J. thanks N. N. Nikolaev and R. Schiavilla for useful comments on
the manuscript.  S.J.  is grateful to the Alexander von Humboldt
Foundation for financial support during part of this work.  This work
was in part supported by funds provided by the U.S. Department of
Energy (D.O.E.)  under cooperative research agreement
\#DF-FC02-94ER40818 and \#DE-AC05-84ER40150.

\appendix
\section{Response Functions and Current Operator}

For the convenience of the reader, we provide the definition of
the response functions and identify the different parts of the
electromagnetic current. For more information on exclusive electron
scattering in general, the reader is referred to \cite{raskintwd}; for
the electromagnetic current operator and the relativistic effects
incorporated in it, see \cite{relcur}.

The differential cross section is equal to

\begin{eqnarray}
\left ( \frac{ d \sigma^5}{d \epsilon' d \Omega_e d \Omega_N}
\right ) _{fi}^h  & = & 
\frac{m_N \, m_f \, p_N}{8 \pi^3 \, m_i} \, \sigma_{Mott} \, 
f_{rec}^{-1} \, \nonumber \\
& & \Big[ \left ( v_L R_{fi}^L +   v_T R^T_{fi}
 + v_{TT} R_{fi}^{TT} + v_{TL} R_{fi}^{TL} \right )
  \nonumber \\
& & +  h \left ( v_{T'} R_{fi}^{T'} +  v_{TL'} R_{fi}^{TL'}
\right ) \Big] \, ,
\end{eqnarray}
where $m_i$, $m_N$ and $m_f$ are the masses of the target nucleus, the
ejectile nucleon and the residual system, $p_N$ and $\Omega_N$ are the
momentum and solid angle of the ejectile, $\epsilon'$ is the energy of
the detected electron and $\Omega_e$ is its solid angle.  The helicity
of the electron is denoted by $h$.  The coefficients $v_K$ are the
leptonic coefficients, and the $R_K$ are the response functions
which are defined by

\begin{eqnarray}
R_{fi}^L & \equiv & | \rho (\vec q)_{fi}|^2 \nonumber \\
R_{fi}^T & \equiv & | J_+ (\vec q)_{fi}|^2 
+ | J_- (\vec q)_{fi}|^2  \nonumber  \\
R_{fi}^{TT} & \equiv &  2 \, \Re \, \big[ J_+^* (\vec q)_{fi} \,
J_- (\vec q)_{fi} \big] \nonumber  \\
R_{fi}^{TL} & \equiv & - 2 \, \Re \, \big[ \rho^* (\vec q)_{fi} \,
( J_+ (\vec q)_{fi} - J_- (\vec q)_{fi}) \big] \nonumber \\
R_{fi}^{T'} & \equiv & | J_+ (\vec q)_{fi}|^2 -
 | J_- (\vec q)_{fi}|^2  \nonumber  \\
R_{fi}^{TL'} & \equiv & - 2 \, \Re \, \big[ \rho^* (\vec q)_{fi} \, 
( J_+ (\vec q)_{fi} + J_- (\vec q)_{fi}) \big] \, , 
\label{defresp}
\end{eqnarray}
where the $J_{\pm}$ are the spherical components of the current.  For
our calculations, we have chosen the following kinematic conditions:
the z-axis is parallel to $\vec q$, the missing momentum is defined as
$\vec p_m \equiv \vec q - \vec p_N$, so that in PWIA, the missing
momentum is equal to the negative initial momentum of the struck
nucleon in the nucleus, $\vec p_m = -\vec p$. We denote the angle
between $\vec p_m$ and $\vec q$ by $\theta$, and the term ``parallel
kinematics'' indicates $\theta = 0^o$, ``perpendicular kinematics''
indicates $\theta = 90^o$, and ``antiparallel kinematics'' indicates
$\theta = 180^o$.  Note that both this definition of the missing
momentum and the definition with the other sign are used in the
literature.  In this paper, we assume that the experimental
conditions are such that the kinetic energy of the outgoing nucleon
and the angles of the missing momentum, $\theta$ and the azimuthal
angle $\phi$, are fixed.  The kinetic energy of the outgoing proton is
fixed to 1 GeV. For changing missing momentum, the transferred energy
and momentum change accordingly.

The electromagnetic current operator
\begin{equation}
J^\mu (P\Lambda ;P^{\prime }\Lambda ^{\prime })=\bar u(P^{\prime }\Lambda
^{\prime })\left[ F_1\gamma ^\mu +\frac i{2m_N}F_2\sigma ^{\mu \nu }Q_\nu
\right] u(P\Lambda )  \label{sn1}
\end{equation}
can be rewritten in a form that is more suitable for application to nuclear
problems:
\begin{equation}
J^\mu (P\Lambda ;P'\Lambda ^{\prime }) \equiv 
\chi _{\Lambda
^{\prime }}^{\dagger }  \, \, \bar{J}^{\mu }(P;P^{\prime }) 
\, \, \chi _{\Lambda }^{{}}
\end{equation}
with
\begin{eqnarray}
\bar J^o &=& \rho = f_o \left (
\xi _o + \, i \, \, \xi _o^{\prime } \, \left( \vec q \times \vec p
\right) \cdot \vec{\sigma }   \right ) \nonumber \\
\bar J^3 &=& \, \,  \frac{\omega}{q} \, \, \bar J^o  \nonumber \\
\bar J^{\bot } &=& f_o \left (  \xi _{1} \left[ \, \vec p
-\left( \frac{\vec q
\cdot \vec p}{q ^2}\right) \vec q \, \right] -i
\left\{ \xi _1^{\prime } \left( \vec q \times \vec{\sigma }
\right) \right.  \right. \nonumber\\
  &+&  \left. \left. \xi _2^{\prime } \left( \vec q \cdot \vec{\sigma }
\right) \left( \vec q \times \vec p \right) +\xi
_3^{\prime }\left[ \left( \vec q \times \vec p \right)
\cdot \vec{\sigma }\right] \left[ \vec p -\left( \frac{\vec q 
\cdot \vec p }{q ^2}\right) \vec q \right]
\right\} \right )  .
\end{eqnarray}
Here, $f_o, \xi_i, \xi_i'$ are all functions of $\omega, q, p^2$;
their explicit forms are given in \cite{relcur}. For the reasons
explained in \cite{relcur}, we refer to the operator associated with
$\xi_o$ as zeroth-order charge operator, we call the term containing
the $\xi_o'$ first-order spin-orbit operator, the term containing
$\xi_1$ first-order convection current, the term containing $\xi_1'$
zeroth-order magnetization current, the term containing $\xi_2'$
first-order convective spin-orbit term, and the term containing
$\xi_3'$ second-order convective spin-orbit term.  In this paper, we
have used the current expanded up to first order in the initial
nucleon's momentum and retained terms of all order in the transferred
energy $\omega$ and the transferred momentum $q$.

Note that we retain more terms in the current than with the commonly
used strict nonrelativistic reduction, which assumes that the
transferred momentum $q$ is smaller than the nucleon mass, and that
both the initial nucleon momentum and the transferred energy are 
smaller than $q$ and therefore much smaller than the nucleon mass. 
Under these assumptions, the current operator simplifies to the form
\begin{eqnarray}
\bar J^o_{nonrel} &=& G_E \nonumber \\
\bar J^{\perp}_{nonrel} &=& - \frac{i}{2 \, m_N} \,  G_M \, \left (
\vec q \times \vec \sigma \right ) + \frac{1}{m_N} \, G_E \,
\left ( \vec p - \frac{\vec q \cdot \vec p}{q^2} \, \vec q
\right ) \nonumber \,,
\end{eqnarray}
which contains only the zeroth-order charge operator, the zeroth-order
magnetization current and the first-order convection current.



\begin{table}
\caption{NN amplitude parameters for the central and spin-orbit terms
from the Glauber proton-nucleus scattering analysis. Parameters are
given in the $NN$ c.m. frame.} \vspace{1cm}

\begin{tabular}{cccccc}
$\sigma_{tot}^{NN}$ / mb  
& $b_o$  / fm
&$\rho$   
& $\gamma$  / fm
& $b_s$   / fm
& $\rho_s$  
\\ [0.5ex]
\hline \hline
41.1 & 0.48 & -0.48 & 0.16 & 0.65 & -0.24   \\
\end{tabular}
\label{tabnng}
\end{table}

\begin{table}
\caption{NN amplitude parameters for the central and spin-orbit terms
from the phase shift analysis analysis. Parameters are given in the
$NN$ c.m. frame.} \vspace{1cm}
\label{tabnnp}
\begin{tabular}{cccccc}
$\sigma_{tot}^{NN}$ / mb  
& $b_o$  / fm
&$\rho$   
& $\gamma$  / fm
& $b_s$   / fm
& $\rho_s$  
\\ [0.5ex]
\hline \hline
36.9 & 0.57 - i 0.07 & -0.47 & 0.20 & 0.53 + i 0.05 & 0.52   \\
\end{tabular}
\end{table}

\begin{table}
\caption{NN amplitude parameters for the spin-orbit terms
from the Glauber proton-nucleus scattering analysis. Parameters are
given in the $NN$ lab frame.} \vspace{1cm}

\begin{tabular}{cccc}
nucleon   
& $\gamma$  / fm
& $b_s$   / fm
& $\rho_s$  
\\ [0.5ex]
\hline \hline
proton  & 0.15 & 0.65 - i 0.03 & -0.44   \\
neutron & 0.17 & 0.64 - i 0.03 & -0.02   \\
\end{tabular}
\label{tablab}
\end{table}

\begin{table}
\caption{NN amplitude parameters for the spin-orbit terms
from the $N N$ phase shift analysis. Parameters are
given in the $NN$ lab frame.} \vspace{1cm}

\begin{tabular}{cccc}
nucleon   
& $\gamma$  / fm
& $b_s$   / fm
& $\rho_s$  
\\ [0.5ex]
\hline \hline
proton  & 0.18 & 0.53 + i 0.07 & 0.41   \\
neutron & 0.21 & 0.52 + i 0.03 & 0.64   \\
\end{tabular}
\label{tablabpsa}
\end{table}

\end{document}